\documentclass[aps,prb,showpacs,twocolumn,superscriptaddress,floatfix]{revtex4}

\usepackage{graphicx}
\usepackage{epstopdf}
\usepackage{times}
\usepackage{amsmath}
\usepackage{amssymb}
\usepackage{amsfonts}
\usepackage{latexsym}

\def\be{\begin{equation}}
\def\ee{\end{equation}}
\def\bea{\begin{eqnarray}}
\def\eea{\end{eqnarray}}
\def\bs{\boldsymbol}
\def\vec{\mathbf}
\def\mc{\mathcal}

\begin{document}

\title{Dzyaloshinskii-Moriya anisotropy and non-magnetic impurities in the 
$s = 1/2$ kagome system ZnCu$_3$(OH)$_6$Cl$_2$} 

\author{Ioannis Rousochatzakis}
\affiliation{Institute of Theoretical Physics (CTMC), \'Ecole  Polytechnique 
F\'ed\'erale de Lausanne, CH-1015 Lausanne, Switzerland}

\author{Salvatore R. Manmana}
\affiliation{Institute of Theoretical Physics (CTMC), \'Ecole  Polytechnique 
F\'ed\'erale de Lausanne, CH-1015 Lausanne, Switzerland}

\author{Andreas M. L\"auchli}
\affiliation{Max-Planck-Institut f\"ur Physik komplexer Systeme,  D-01187 
Dresden, Germany}

\author{Bruce Normand}
\affiliation{Institute of Theoretical Physics (CTMC), \'Ecole Polytechnique 
F\'ed\'erale de Lausanne, CH-1015 Lausanne, Switzerland}

\author{Fr\'ed\'eric Mila}
\affiliation{Institute of Theoretical Physics (CTMC), \'Ecole Polytechnique 
F\'ed\'erale de Lausanne, CH-1015 Lausanne, Switzerland}

\date{\today}


\begin{abstract}
Motivated by recent nuclear magnetic resonance experiments on 
ZnCu$_3$(OH)$_6$Cl$_2$, we present an exact-diagonalization study of the 
combined effects of non-magnetic impurities and Dzyaloshinskii-Moriya (DM) 
interactions in the $s = 1/2$ kagome antiferromagnet. The local response to 
an applied field and correlation-matrix data reveal that the dimer freezing 
which occurs around each impurity for $D = 0$ persists at least up to 
$D/J\simeq 0.06$, where $J$ and $D$ denote respectively the exchange and 
DM interaction energies. The phase transition to the ($Q = 0$) semiclassical, 
120$^\circ$ state favored at large $D$ takes place at $D/J\simeq 0.1$. 
However, the dimers next to the impurity sites remain strong up to values 
$D \sim J$, far above this critical point, and thus do not participate 
fully in the ordered state. We discuss the implications of our results 
for experiments on ZnCu$_3$(OH)$_6$Cl$_2$.
\end{abstract}

\maketitle

\section{Introduction}

The $s = 1/2$ antiferromagnetic (AFM) Heisenberg model on the two-dimensional 
kagome lattice is one of the simplest models in frustrated quantum magnetism, 
but displays some of the most complex behavior known.\cite{review} As such, 
it has for decades maintained its position at the forefront in the search 
for novel quantum-mechanical phases of matter, such as the 
resonating-valence-bond (RVB) spin-liquid state proposed initially by 
Anderson.\cite{Anderson} The nature of the ground state of the $s = 1/2$ 
kagome AFM has still not been fully established: the variety of competing 
phases proposed in the literature includes valence-bond crystals 
(VBCs),\cite{VBS} gapped spin liquids,\cite{SL1,SL2,SL3} and gapless 
critical phases.\cite{CP1,CP2} Under these circumstances, the discovery 
that the kagome AFM is extremely sensitive even to the smallest perturbations, 
such as the presence of anisotropies\cite{Elhajal,Ballou} or of non-magnetic 
impurities,\cite{Dommange,Lauchli} is completely consistent.   

In this context, the recent discovery\cite{Shores} of the mineral 
herbertsmithite, ZnCu$_3$(OH)$_6$Cl$_2$, has attracted very considerable 
attention, because it represents a structurally perfect realization of 
the $s = 1/2$ kagome AFM. This material is composed of Cu$^{2+}$ ions 
($s = 1/2$) arranged on kagome planes separated by triangular layers 
of non-magnetic Zn$^{2+}$ ions. Despite the large AFM exchange, $J \simeq 
170$-190 K in this material,\cite{Rigol,Misguich} in experiment there is 
no evidence of long-ranged magnetic order or even of spin freezing at any 
temperatures down to 50 mK.\cite{muons,Helton} There is in addition no 
sign of a spin gap in the excitation spectrum.\cite{Helton,Olariu} 

However, it has also been established that in ZnCu$_3$(OH)$_6$Cl$_2$ there 
is a significant (5-10\%) intersite exchange of Cu$^{2+}$ ions with the 
Zn$^{2+}$ ions intended to separate the kagome planes.\cite{impurities1, 
impurities2} The displaced Cu$^{2+}$ ions are thought to account for the 
large Curie tails observed in powder susceptibility measurements\cite{Bert, 
Helton} and for the field-dependent, Schottky-like anomaly in the specific 
heat.\cite{impurities2} At the same time, the Zn$^{2+}$ ions displaced 
into the kagome planes play the role of non-magnetic vacancies, which are 
known\cite{Dommange,Lauchli} to modify the ground state properties in a 
nontrivial way. However, these impurities can also be considered as a probe 
of kagome physics, and in this respect the recent $^{17}$O nuclear magnetic 
resonance (NMR) experiments reported by Olariu {\it et al.}\cite{Olariu} 
offer extensive insight. While each O$^{2-}$ ion is coupled predominantly 
to two neighboring Cu$^{2+}$ ions in the kagome planes, the NMR spectra 
revealed a broad distribution of local susceptibilities. The relative 1:2
ratio between the line shifts of the two leading features in the $^{17}$O 
spectrum was explained on the basis of distinguishing between two groups 
of O sites in the doped kagome planes, those [``Defect'' (D)] probing the 
magnetic polarization directly next to an impurity site, and hence coupled 
only to one Cu spin, and those [``Main'' (M)] reflecting the polarizations 
of all other sites. This interpretation was also consistent with the 
relative intensities of the two lines based on the expected in-plane 
dopant concentration of approximately 5\%. Another major finding of these 
experiments was the observation that the (M) and (D) line shifts approach 
a finite value as $T \to 0$, suggesting a non-singlet ground state without 
a gap. This result can also be inferred from the non-activated behavior 
of the nuclear spin-lattice relaxation rate, $1/T_1$, at low 
temperatures.\cite{Olariu}

While the true nature of the non-magnetic ground state of the pure, 
$s = 1/2$ Heisenberg kagome AFM remains unknown, all of the guidance 
obtained from VBC or RVB constructions suggests a total singlet. Indeed, 
the nearest-neighbor RVB basis has been shown to deliver a semi-quantitative 
account of some of the properties of the system, both without\cite{SL2,rmm} 
and with\cite{Dommange} impurities, and for this reason a non-singlet ground 
state presents a considerable challenge. However, by the nature of its 
structure, a triangular geometry involving active $d_{x^2-y^2}$ and 
$d_{3z^2-r^2}$ orbitals, the Cu--Cu bonds in ZnCu$_3$(OH)$_6$Cl$_2$ are 
not centrosymmetric. The resulting Dzyaloshinskii-Moriya (DM) 
interactions\cite{DM} induce quite generally a small admixture of triplet 
states into a singlet ground state,\cite{Rigol} and thus have been 
discussed\cite{Olariu} as a likely explanation for the finite response 
observed at zero temperature. The DM interactions in herbertsmithite have
been determined recently by electron spin resonance experiments.\cite{Zorko} 
These measurements show that the dominant component of the DM interaction is 
that perpendicular to the kagome plane, which is of order $D \simeq 15$ K 
$\sim 0.08 J$, while the in-plane component, $D'$, although not excluded 
by symmetry, is much smaller ($D'\simeq 2$ K $\sim 0.01 J$).  

Clearly, a complete theoretical description for the ground-state properties 
of ZnCu$_3$(OH)$_6$Cl$_2$, including the magnetization response around the 
impurity sites, must take into account the combined effect of both the DM 
interactions and the non-magnetic impurities in the kagome planes. Among 
the many approaches adopted to gain further insight into the kagome system, 
it is well known that the clean, classical kagome AFM without DM anisotropy 
has an extensive ground-state degeneracy,\cite{Moessner_Review_Class_Deg} 
which is fully lifted by quantum fluctuations only beyond harmonic 
order.\cite{HarrisKallinBerlinsky,Chubukov,ChanHenley} The inclusion of 
DM anisotropy perpendicular to the kagome plane selects immediately the 
uniform ($Q = 0$), three-sublattice state.\cite{Elhajal,Ballou} C\'epas 
{\it et al.} have shown recently\cite{Cepas} that in the quantum ($s = 1/2$) 
kagome AFM, this ordered phase can be stabilized only for $D/J \gtrsim 0.1$. 
This implies that herbertsmithite may well be very close to the critical 
point where quantum fluctuation effects promote a disordered ground state. 
The linear susceptibility of the clean kagome AFM in the presence of DM 
interactions has been analyzed\cite{Tovar} using a perturbative expansion 
about a short-ranged VBC ground-state scenario, while the effect of 
non-magnetic impurities without DM interactions has been studied in 
Refs.~[\onlinecite{Dommange,Lauchli,Gregor,Rozenberg}]. 
It has been shown that one of the central consequences of a non-magnetic 
impurity in the quantum kagome AFM is a characteristic dimer freezing, 
which takes place around the impurity site due to frustration relief in 
the affected triangles.\cite{Dommange,Lauchli} 

Here we present an extensive exact-diagonalization study which includes 
the effects of both DM anisotropy and non-magnetic impurities. Our results 
can be summarized as follows. For $D/J \lesssim 0.06$, the effect of 
frustration relief around the impurity\cite{Dommange,Lauchli} causes the 
formation of strong, local dimers whose magnetization response, a staggered 
moment directed along $\vec{D} \times \vec{B}$, is qualitatively similar to 
that of isolated dimers. In this regime we find surprisingly large variations 
in the magnitude of the response on the different induced dimers. For sites 
far away from the impurity, we argue that the response must be uniform and 
much smaller in magnitude. For $D/J \gtrsim 0.1$, the system enters the 
$Q = 0$, 120$^\circ$, semiclassically ordered state,\cite{Elhajal,Ballou, 
Cepas} and the effect of impurities becomes very short-ranged. However, 
even here we find that the dimers next to the impurity remain strong up 
to values $D/J\sim 1$, meaning that their spins remain correlated with 
each other, rather than participating in the ordered state, for $D/J 
\lesssim 1$.

\begin{figure}[!b]
\includegraphics[width=0.45\textwidth]{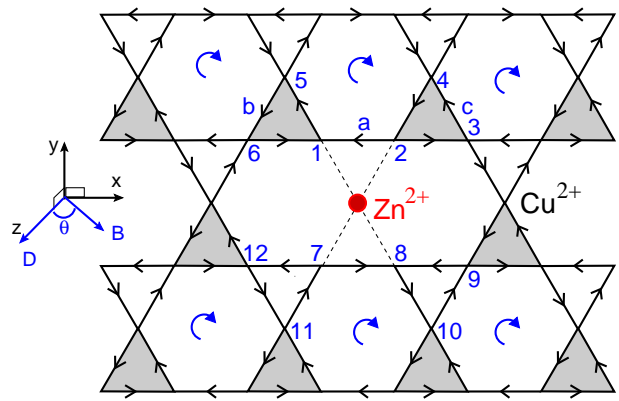}
\caption{(Color online) The kagome lattice of Cu$^{2+}$ ions ($s = 1/2$), 
containing a single, non-magnetic Zn$^{2+}$ impurity. Bond arrows 
define site labels $i$ and $j$ in each interaction term $\vec{D} \cdot 
\vec{s}_i \times \vec{s}_j$, while $\vec{D} = D \vec{e}_z$ with $D > 0$ 
for all bonds. Curving arrows denote the $\mathsf{C}_6$ symmetry about 
the hexagon centers.}
\label{fig:system}
\end{figure}

This article is organized as follows. In Sec.~\ref{sec:model} we 
introduce the model Hamiltonian and discuss some of the symmetry 
aspects of the clusters considered in our exact-diagonalization 
calculations. The results for the local magnetization response 
are presented and analyzed in Sec.~\ref{sec:local}. In 
Sec.~\ref{sec:correlations}, we investigate the dominant magnetic 
correlations on the basis of the full correlation matrix (the ``natural 
orbital'' method). We conclude in Sec.~\ref{sec:conclusions} with a 
discussion of the implications of our results for experimental measurements 
on ZnCu$_3$(OH)$_6$Cl$_2$. We also include two appendices which elaborate 
on the magnetization response of a minimal (four-site) cluster 
(App.~\ref{app:4sites}) and on the natural orbital method 
(App.~\ref{app:Cij}).  

\section{Model}\label{sec:model}

We consider a spin-1/2 model on the kagome lattice with a single, 
non-magnetic impurity described by the Hamiltonian
\begin{equation}
\mathcal{H} = J \sum\limits_{\langle ij \rangle} \vec{s}_i \cdot \vec{s}_j
 + \sum\limits_{\langle ij \rangle} \vec{D}_{ij} \cdot \left( \vec{s}_i 
\times \vec{s}_j \right) - \vec{B} \cdot \vec{S},
\label{eq:hamiltonian}
\end{equation}
with periodic boundary conditions. The first term is the Heisenberg 
exchange energy between nearest-neighbor spins $\langle ij \rangle$, the 
second term represents the DM interactions, and the last term 
is the Zeeman energy of the total spin $\vec{S} = \sum_i \vec{s}_i$ in a 
field $\vec{B}$. In what follows, we work in the fixed reference frame 
$xyz$ shown in Fig.~\ref{fig:system}. The field is taken to be in the 
$xz$-plane at an angle $\theta$ from the $z$-axis. The DM vectors on every 
bond are taken to be perpendicular to the kagome plane and are fixed by the 
symmetry of the clean system, namely translations and $\mathsf{C}_6$ 
rotations around the hexagon centers. We have chosen the site-labelling 
convention, denoted in Fig.~\ref{fig:system} by the directionality of the 
arrows from $i$ to $j$, such that $\vec{D}_{ij} = D\vec{e}_z$ with $D > 0$ 
for all bonds. In this study we do not consider an in-plane DM component, 
because in ZnCu$_3$(OH)$_6$Cl$_2$ this is known\cite{Zorko} to be much 
weaker than the out-of-plane component. We note, however, that the in-plane 
DM problem on the kagome lattice is quite significantly different from its 
out-of-plane counterpart studied here.\cite{Elhajal}

We will focus on the magnetization response of the system in finite 
fields, up to the values $B \sim 10$ T ($B \lesssim J/20$) probed by 
experiment. For a general field orientation $\theta$, the $\mathsf{U}(1)$ 
spin rotation symmetry around the $z$-axis is broken and thus the total 
magnetization $S^z$ is not a good quantum number. Because of the impurity 
site, the same is true also for the momentum. Thus we can treat kagome 
clusters with a maximum of $N = 26$ magnetic sites, and here we show 
primarily the results for $N=14, 20$, and $26$. These clusters are 
symmetric with respect to spatial inversion through the impurity site.
For a non-degenerate ground state, this sets the constraints 
\be
\langle \vec{s}_1\rangle=\langle \vec{s}_{8}\rangle, ~\langle \vec{s}_2 
\rangle=\langle \vec{s}_{7}\rangle,~\langle \vec{s}_3\rangle=\langle 
\vec{s}_{12}\rangle,~\dots
\ee
Furthermore, the clusters with $N = 14$ and $26$ are also spatially 
symmetric under reflection through the $xz$-plane which passes through 
the impurity site (Fig.~\ref{fig:system}), whence their Hamiltonian 
(\ref{eq:hamiltonian}) remains invariant under the corresponding mirror 
operation (spatial reflection followed by the time reversal). For a 
non-degenerate ground state, this gives in addition 
\bea
\langle s_1^{x,z}\rangle &=& \langle s_{7}^{x,z}\rangle,~\langle s_1^y 
\rangle = -\langle s_{7}^y\rangle \nonumber\\
\langle s_2^{x,z}\rangle &=& \langle s_{8}^{x,z}\rangle,~\langle s_2^y 
\rangle = -\langle s_{8}^y\rangle,~\dots 
\label{eq:symm2}
\eea
This mirror symmetry is also present in the clean, infinite kagome 
lattice. Thus if the vacant site were occupied by a spin $\vec{s}_0$,
Eq.~(\ref{eq:symm2}) would require $\langle s_0^y \rangle = 0$. 
The remaining symmetries of the clean kagome lattice would then enforce 
$\langle s_i^y \rangle = 0$ for all sites $i$, an argument we will employ 
below to demonstrate that in the disordered phase ($D/J\lesssim 0.1$)
there can be no staggered magnetic response along the $y$-axis sufficiently 
far from a vacant site. 

\begin{figure}[!t]
\includegraphics[width=0.45\textwidth]{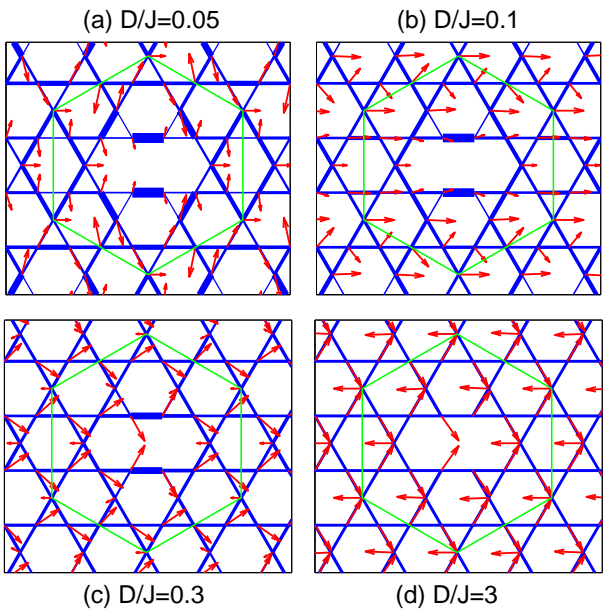}
\caption{(Color online) In-plane magnetizations and bond spin correlation 
functions for the 26-site kagome cluster with $B = J/20$ and $\theta = 
30^\circ$. The lengths of the arrows are proportional to $\left( \langle 
s_i^x \rangle^2 + \langle s_i^y \rangle^2 \right)^{1/2}$ and the 
thicknesses of the bonds to $\langle \vec{s}_i \cdot \vec{s}_j \rangle$.}
\label{fig:profiles}
\end{figure} 

In addition to these ``symmetric'' clusters, we have also investigated 
clusters with odd numbers of magnetic sites, particularly with $N = 17$ 
and $23$, whose ground state has a finite moment $S_z = 1/2$. As shown 
in Ref.~[\onlinecite{Didier}], this moment does not form a bound state 
around the impurity site, as is the case in systems such as the 
checkerboard lattice,\cite{Didier} but is delocalized over the entire 
cluster. Thus we expect no difference between clusters with even or odd 
numbers $N$ of magnetic sites in the thermodynamic limit, and our results 
indeed conform with this picture. 

\section{Local response}\label{sec:local}

\begin{figure}[!t]
\includegraphics[width=0.48\textwidth]{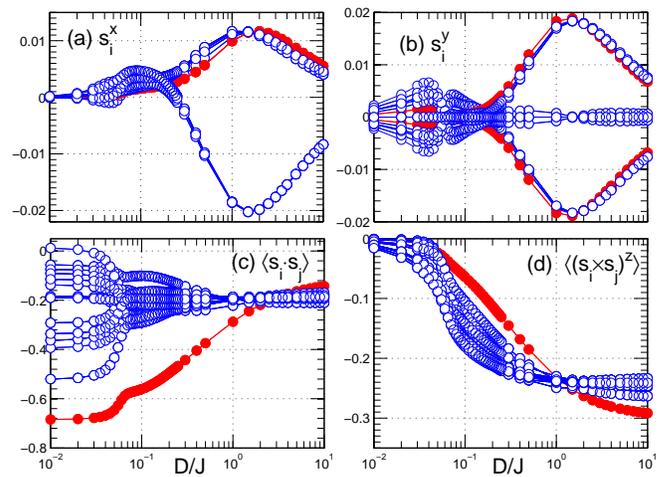}
\caption{(Color online) $D$-dependence of (a) $\langle s_i^x \rangle$, 
(b) $\langle s_i^y \rangle$, (c) $\langle \vec{s}_i \cdot \vec{s}_j \rangle$,
and (d) $\langle \vec{s}_i \times \vec{s}_j \rangle^z$ for the 26-site kagome 
cluster with $B = J/20$ and $\theta = 30^\circ$. $i$ and $j$ in panels (c) 
and (d) are nearest-neighbor sites only. The curves which correspond to the 
sites or bonds closest to the impurity are designated by (red) filled symbols.}
\label{fig:all26sites}
\end{figure}

Figure \ref{fig:profiles} shows the local in-plane magnetizations $\langle 
s_i^{x,y} \rangle$ (red arrows) and the local spin correlation functions 
$\langle \vec{s}_i \cdot \vec{s}_j\rangle$ (described quantitatively by 
the thickness of each bond $ij$) for the 26-site cluster with $B = J/20$, 
$\theta = 30^\circ$, and for four representative values of $D/J$. The 
full $D/J$-dependence of the in-plane magnetizations, the correlation 
amplitudes, and the local twist amplitudes $\langle (\vec{s}_i \times 
\vec{s}_j)^z\rangle$ is shown for each type of site or bond in 
Fig.~\ref{fig:all26sites}. These results demonstrate that there exist 
two primary regimes with qualitatively different magnetic response, 
whose properties we discuss in detail in the following subsections. In 
summary, these are (i) the small-$D/J$ regime ($D/J\lesssim 0.06$), where 
there is a local, dimer-like response for the sites around the impurity, 
and (ii) the large-$D/J$ regime ($D/J\gtrsim 0.1$), which is characterized 
by the expected\cite{Elhajal,Ballou,Cepas} in-plane, 120$^\circ$ magnetic 
ordering pattern, albeit with a special response from the sites right next 
to the vacancy. We also find an apparent intermediate regime ($0.06 
\lesssim D/J \lesssim 0.1$) with a peculiar magnetization pattern around 
the impurity, but with no evidence of long-ranged magnetic correlations, 
and we will also characterize this to the extent that our calculations allow.

\subsection{$D \lesssim 0.06J$: local dimer-like response}

We begin by considering the behavior of the nearest-neighbor spin 
correlations, $\langle \vec{s}_i \cdot \vec{s}_j\rangle$, which are shown 
in Fig.~\ref{fig:all26sites}(c) and are also denoted by the bond thicknesses 
in Fig.~\ref{fig:profiles}. A low-$D$ regime, where the correlations 
are controlled by the exchange interaction ($J$) and by the presence of the 
impurity, can be seen to exist for $D/J \lesssim 0.06$, although in fact 
the correlation amplitudes begin to deviate from their zero-$D$ limit at 
$D/J \sim 0.03$. This limit was studied in detail 
in Ref.~[\onlinecite{Dommange}] and is characterized by a large and 
oscillatory variation of the correlation amplitudes as a function of 
distance from the impurity. As shown in Fig.~\ref{fig:profiles}(a), the 
strongest dimerization is found on the bonds nearest the vacancy, where 
$\langle \vec{s}_i \cdot \vec{s}_j \rangle \simeq -0.69$ is very close to  
the value of $-0.75$ expected for an isolated dimer. This local ``dimer 
freezing'' is a key consequence of the frustration relief in the two 
triangles containing the vacancy\cite{Dommange,Lauchli} and, as we will 
discuss below, has direct implications for the magnetization response of 
the corresponding sites in the small-$D/J$ regime.

The local twist amplitudes $\langle (\vec{s}_i \times \vec{s}_j)^z 
\rangle$ shown in Fig.~\ref{fig:all26sites}(d) represent, from 
Eq.~(\ref{eq:hamiltonian}), the DM contribution to the total magnetic 
energy. As expected for the small $D/J$ regime, these scale linearly 
with $D/J$ for all bonds, although with a quite inhomogeneous distribution 
of slopes depending on the distance of the bond from the impurity. In 
particular, the bonds closest to the impurity have one of the smallest 
twist amplitudes, as might be anticipated from their strong dimerization.

Turning next to the magnetization response shown in 
Figs.~\ref{fig:profiles}(a) and \ref{fig:all26sites}(a)-(b), this develops 
predominantly in the $xy$-plane, as expected given the easy-plane character 
of the DM anisotropy, and the induced spin components are of order 
$\langle s_i^{x,y} \rangle \sim 10^{-3}$--$10^{-2}$. By contrast, the 
out-of-plane response is very small ($\langle s_i^z \rangle \lesssim 
10^{-6}$), and thus will not be considered in the following. With the 
exception of sites strongly influenced by the periodic boundary conditions, 
the magnetization response shows a number of features which are also present 
in the 20-site cluster and can be understood qualitatively on the basis of 
the strong local dimer formation around the impurity. Indeed, the direction 
of the induced magnetizations at the strongly dimerized bonds is qualitatively 
similar to the response expected\cite{Miyahara} if they were isolated: a 
staggered polarization in the direction $\vec{D} \times \vec{B}$ (here the 
$y$-axis) and a uniform polarization in the direction $(\vec{D} \times 
\vec{B}) \times \vec{D}$ ($x$-axis). At the quantitative level, the actual 
magnitude of the magnetization response goes beyond the isolated-dimer 
picture, showing the striking feature that the moments induced on the four 
``next-most-strongly'' dimerized bonds [bond (3,4) and its equivalents in 
Fig.~\ref{fig:system}] are larger by a factor of 3-4 than on the bonds 
closest to the impurity. The origin of this result, which we probe in 
detail below, is rather deeper than a simple anticorrelation between 
spin correlation amplitude and induced moment. It lies in the fact that, 
while the ground-state wave function close to the impurity can be described 
as an approximate product state of local dimers, the actual nature of the 
excitations may nevertheless differ significantly from such a picture. 

\begin{figure}[!t]
\includegraphics[width=0.4\textwidth]{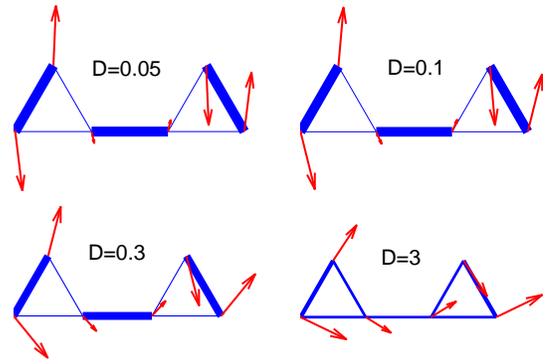}
\caption{(Color online) In-plane magnetization pattern (red arrows) and 
bond strengths (bond thickness in blue) on the isolated six-site cluster 
discussed in Sec.~\ref{sec:local}, for $B=J/20$ and $\theta=30^\circ$.}
\label{fig:6sites}
\end{figure}

To clarify this point, we isolate the sites 1-6 of Fig.~\ref{fig:system} 
and consider their response in the small-$D/J$ regime. Figure \ref{fig:6sites} 
shows the exact magnetization response of this cluster with parameters $B = 
J/20$, $\theta = 30^\circ$, and for the same field orientation and four 
representative values of $D/J$ shown in Fig.~\ref{fig:profiles}. The 
response for $D = 0.05J$ is very similar to that of the sites 1-6 in 
Fig.~\ref{fig:profiles}, and in fact the same physics can also be 
shown to dominate the response of a four-site cluster (sites 1-4 of 
Fig.~\ref{fig:system}), as detailed in App.~\ref{app:4sites}. In all 
cases, the magnetizations induced on bond (3,4) are larger than those on 
bond (1,2) by a fixed factor of up to 5. The qualitative interpretation 
(App.~\ref{app:4sites}) of this result lies in the different character 
of the local excitations on bonds (1,2) and (3,4): a triplet excitation 
on bond (1,2) is a true eigenstate of the isotropic Hamiltonian, whereas 
a similar triplet excited on the bond (3,4) is not, and may propagate away. 
As a result, the magnetization response of bond (1,2) is to first order in 
$D/J$ the same as that of an isolated dimer, while that of bond (3,4) 
will be almost 5 times larger in the extreme case of the four-site 
cluster. The same arguments apply not only on the six-site cluster, but 
also to the impure kagome lattice, although in this latter case a triplet 
excitation on the bonds next to the impurity does not remain fully 
localized because the bonds are not fully ``frozen.'' More generally, 
the qualitative agreement between the response of the smallest clusters 
and the response of the corresponding sites in the $20$ and $26$-site 
calculations shows that, in the regime of small $D/J$, the magnetization 
profile around the impurity is governed almost entirely by local correlations. 

The magnetization response far from the impurity is expected to be that 
of the clean kagome lattice with DM interactions. As explained above, the 
staggered magnetization must vanish by symmetry, while a uniform component 
(along the $x$-axis) can survive. This latter scales as $\vec{D} \times 
(\vec{D} \times \vec{B})$ (see also Ref.~[\onlinecite{Tovar}]) and thus 
is very much weaker than the staggered response in close proximity to the 
impurity. This result is corroborated by exact-diagonalization calculations 
on a clean, 12-site kagome cluster (not shown here), which reveal a very 
weak uniform response ($\langle s_i^x \rangle \sim 10^{-7}$). We comment 
that results for larger clean clusters ($N = $18 and 24) suffer from severe 
finite-size effects due to the presence of inequivalent loops in these 
geometries, and thus cannot be used for a quantitative determination 
of the magnetization response far from the impurity in the thermodynamic 
limit.

\subsection{$D \gtrsim 0.1J$: semiclassical, $Q=0$, $120^\circ$ ordered state}

At large values of $D/J$ [Fig.~\ref{fig:profiles}(d)], the magnetization 
profiles correspond closely to the expected\cite{Elhajal,Ballou,Cepas} 
$Q = 0$, semiclassical, three-sublattice (120$^\circ$) state. We note 
here that in this case the magnetic field and the impurity combine to 
pin one of the (infinitely many) degenerate semiclassical states. This 
can be understood physically from the fact that the selected configuration 
is the one in which the spin missing at the impurity site would be 
antiparallel to the field, and thus would contribute a positive Zeeman 
energy. However, Fig.~\ref{fig:profiles}(d) also shows that the spins 
closest to the impurity are slightly tilted away from a perfect 120$^\circ$ 
orientation. In fact these spins do not participate in the ordered state in 
the same way as all the other (``bulk'') spins do, as can be inferred from
the spin correlation amplitudes in Fig.~\ref{fig:all26sites}(c). Above 
$D/J \sim 0.06$, and especially beyond $D/J \sim 0.6$, all the bonds except 
the two closest to the impurity attain very similar strengths. By contrast, 
the dimers next to the impurity (denoted by filled symbols) resist this 
tendency, their spins remaining strongly correlated up to surprisingly 
high values in excess of $D\sim J$. There is no such feature on the 
neighboring bonds, i.e.~there is no indication of a length scale in this 
behavior. This is another primary consequence of frustration relief on 
the triangles hosting the impurity, and will be further corroborated by 
our analysis of the in-plane correlation data in Sec.~\ref{sec:correlations}.

One further comment regarding the behavior of the twist amplitude at large 
$D/J$ is in order here. In the clean kagome system, the twist amplitude is 
the same on every bond, but according to Fig.~\ref{fig:all26sites}(d), the 
sites next to the impurity have a twist amplitude, $\langle( \vec{s} \times 
\vec{s}_j)^z \rangle \simeq -0.3$ for $D/J \to \infty$, larger than on any 
of the other bonds. This reflects the fact that the relative orientation of 
these two spin pairs lies between the 120$^\circ$ orientation of the spins 
on triangles and the 90$^\circ$ orientation favored in an isolated dimer, 
and as such shows the consequence of frustration relief for the DM interaction.

\subsection{Intermediate regime: $0.06\lesssim D/J\lesssim 0.1$} 

Our results for the local magnetic properties (Figs.~\ref{fig:profiles} and 
\ref{fig:all26sites}) imply the presence of an intermediate regime, $0.06 
\lesssim D/J \lesssim 0.1$, which cannot be classed as being in either of 
the above limits. Here the spins develop a peculiar magnetization pattern 
around the impurity site, seemingly indicative of spins with no effective 
interaction which only follow the applied field. This is shown for the case 
$D/J = 0.1$ in Fig.~\ref{fig:profiles}(c), but the analogous picture for 
$D/J = 0.07$ is practically indistinguishable. We have investigated 
this regime in considerable detail as a function of $D/J$, of the applied 
field $B/J$, and of the field angle $\theta$, finding essentially perfect 
linear response to the field down to the lowest energy scales. As we will 
show in Sec.~\ref{sec:correlations}, the in-plane magnetic correlation data 
show no evidence for long-ranged magnetic order in this regime. 

While we cannot exclude the possibility of some type of exotic order in 
this regime, such as a chiral or a spin-nematic phase, we are unable to 
find any evidence of a local order parameter. Thus we feel that, pending 
further studies by other approaches, the most likely explanation for the 
peculiar magnetization patterns we observe lies in finite-size effects. 
One of the reasons that these effects are strong here is the fact that 
a uniform magnetic field is not the appropriate conjugate field for the 
three-sublattice ordered state, and thus competes with this type of order 
close to the critical point where the size of the ordered moment is small. 

\section{Magnetic correlations}\label{sec:correlations} 

Although the analysis of the local magnetization data in Sec.~III 
reveals several major, qualitative features of the ground state of 
Eq.~(\ref{eq:hamiltonian}), a number of issues remain unresolved.
One is whether there exists any type of magnetic order in the window 
$0.06 \lesssim D/J \lesssim 0.1$, and what is the actual critical value 
of $D/J$ where the 120$^\circ$ phase is established. Another concerns the 
fact that the magnetization patterns shown in Fig.~\ref{fig:profiles} give 
few indications regarding the consequences of finite-size effects in our 
calculations, which may be quite different in different parameter regimes, 
and a method is required by which to make this more systematic. Further, 
the actual magnitudes of the moments in the ordered phase cannot be 
inferred from the local magnetization data, because the latter do not 
represent the thermodynamic limit. While we will obtain these 
moments, here we will not pursue the observation, from our numerical data 
for the field-dependence of the magnetizations at large $D$ ($D/J \gtrsim 1$), 
that a very large field ($B/J \sim 1$) is required to establish their full 
lengths.\cite{Capponi}

It is thus necessary to analyze the magnetic correlations in the ground 
state $|\Psi\rangle$ of Eq.~(\ref{eq:hamiltonian}) as a function of $D/J$. 
The easy-plane character of the DM anisotropy allows us to focus on 
in-plane correlations, but the breaking of translational invariance by 
the impurity makes it necessary to consider the full set of these, which 
is contained in an $N \times N$ matrix with entries
\be\label{eq:Cij}
C_{ij} = \langle \Psi| s_i^+s_j^-|\Psi\rangle.
\ee
We remark here that an analysis based on the connected (or cumulant) 
correlation matrix instead of on $\vec{C}$ (\ref{eq:Cij}) gives the same 
results because, for the fields considered in this study, the local 
magnetic moments $\langle s_i^{\pm} \rangle$ are negligibly small compared 
to the connected portion of the correlations, a result which is true for 
both small and large values of $D/J$. A further consequence of this 
is that our low-field correlation data ($B = J/20$ in the calculations to 
follow) are essentially identical to the $B = 0$ case.

For translationally invariant systems, the correlation matrix can be 
partially or fully diagonalized by a simple Fourier transformation. 
In the present case, where translational invariance is absent due to the 
non-magnetic impurity, $\vec{C}$ must be diagonalized numerically. 
The information contained in $\vec{C}$ is very useful, especially if the 
ground state contains long-ranged magnetic correlations in the $xy$-plane. 
In this case, it is known (Refs.~[\onlinecite{Penrose,Yang}] and 
App.~\ref{app:Cij}) that the maximum eigenvalue $\lambda_m$ of $\vec{C}$ 
is macroscopically large, i.e.~$\lambda_m \propto N$. The corresponding 
eigenvector, $\bs{v}_m$, then describes the magnetic mode of the system 
with the dominant fluctuations, and thus is related directly to the spatial 
dependence of the relevant order parameter. In the case at hand, the 
dominant mode $\bs{v}_m$ is not simply a periodic modulation of the spins, 
but will contain valuable information for the non-trivial magnetization 
profile around the impurity. In what follows we exploit this information.

\begin{figure}[t]
\includegraphics[width=0.45\textwidth]{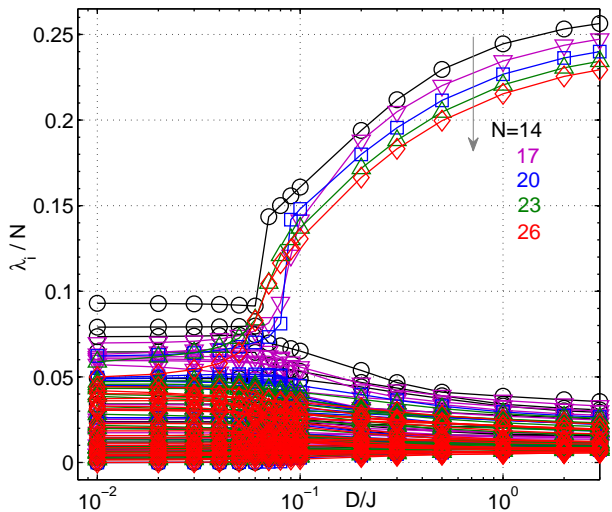}
\caption{(Color online) Eigenvalues $\lambda_i$ of the correlation matrix 
(divided by $N$) as a function of $D/J$ for $N = $14 (circles), 17 (down 
triangles), 20 (squares), 23 (up triangles), and 26 (diamonds). Here $B = 
J/20$ and $\theta=30^\circ$, but the results for $B = 0$ are identical.} 
\label{fig:Cevals}
\end{figure}

\begin{figure}[t]
\includegraphics[width=0.45\textwidth]{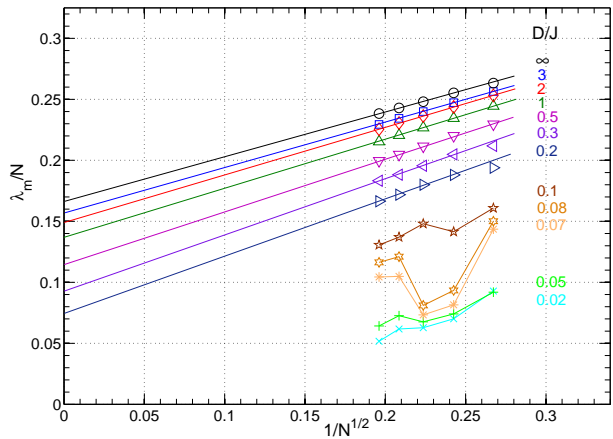}
\caption{(Color online) Scaling with system size of the largest eigenvalue 
$\lambda_m$ (divided by $N$) of the correlation matrix. The solid lines 
correspond to the expected $1/\sqrt{N}$ scaling of the leading corrections 
to the thermodynamic limit in the 120$^\circ$ ordered phase.}
\label{fig:scaling}
\end{figure}

Figure \ref{fig:Cevals} shows the full eigenvalue spectrum of $\vec{C}$ 
for clusters with N$=$14, 17, 20, 23, and 26. The results for $B = 0$ and 
$B = J/20$ are indistinguishable. An essential feature of Fig.~\ref{fig:Cevals} is 
that, as $D/J$ is increased in the region beyond $0.06$, the maximum 
eigenvalue on each of the clusters becomes proportional to $N$, and thus 
very much larger than the remaining eigenvalues; we caution that the 
actual value of this crossover cannot be inferred from the data of 
Fig.~\ref{fig:Cevals}, where it is evident that the curves are still 
some way from the thermodynamic limit, and address this point below. As 
described in App.~\ref{app:Cij}, this means that the system has developed 
long-ranged in-plane magnetic correlations in the regime of large $D/J$. 
For small $D/J$, all the eigenvalues depend only weakly on $D/J$ and are 
closely spaced in magnitude, which is a sign of short-range correlations 
dictated not by $D$ but by $J$.

The finite-size scaling of the dominant eigenvalue $\lambda_m$, normalized 
by $N$, is given in Fig.~\ref{fig:scaling}. The extrapolated values of 
this quantity represent the square of the average magnetic moment in the 
thermodynamic limit, where the spins reach approximately $80\%$ of 
their full moment as $D/J\to \infty$.\cite{factor2} For all values $D/J
 > 0.1$, it is clear that the finite-size corrections scale as $1/\sqrt{N}$, 
as expected for a state of broken U(1) symmetry.\cite{scaling} This scaling 
procedure represents the appropriate means of deducing the existence of 
long-ranged magnetic order, by continuing the curves of Fig.~\ref{fig:Cevals} 
to the infinite-system limit. However, this powerful method shows no 
indication of such order in the regime $0.06 \lesssim D/J \lesssim 0.1$, 
specifying that the transition to the semiclassical state should be taken 
as $D/J \simeq 0.1$. 

\begin{figure}[!b]
\includegraphics[width=0.5\textwidth]{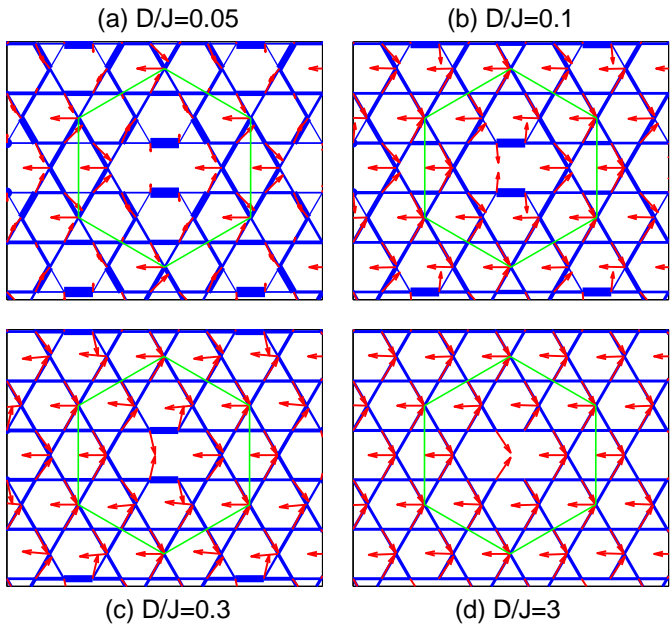}
\caption{(Color online) Magnetization profile for the 26-site kagome cluster 
in zero field, corresponding to the eigenvector $\bs{v}_m$ of $\vec{C}$ with 
the largest eigenvalue $\lambda_m$. The in-plane moments are given by the 
real and imaginary parts of $\bs{v}_m$. Note that this mode is unique up to 
a global U(1) rotation which is related to the arbitrary phase of $\bs{v}_m$.}
\label{fig:CorrDomMode}
\end{figure}

The magnetization profile corresponding to the dominant eigenmode 
$\bs{v}_m$ also contains important information, which is shown in 
Fig.~\ref{fig:CorrDomMode} for the four representative values of $D/J$ and 
represented by two-dimensional arrows whose components are the real and 
imaginary parts of $\bs{v}_m$. At $D/J\lesssim 0.06$, there is no dominant 
mode as is the case at large $D$, but the strongest mode shown in 
Fig.~\ref{fig:CorrDomMode}(a) corresponds nevertheless to the pattern of 
strong spin correlations around the impurity [Fig.~\ref{fig:profiles}(a)]: 
the correlations in this mode are confined to the strong bonds next to the 
impurity, where the spins are almost antiparallel. The strength of these 
local correlations is governed by $J$, which is the reason why $\lambda_m$ 
remains essentially $D$-independent for $D/J\lesssim 0.06$ in 
Fig.~\ref{fig:Cevals}. We emphasize again that the profile shown in 
Fig.~\ref{fig:CorrDomMode}(a) does not represent the actual magnetization 
response -- this is shown in Fig.~\ref{fig:profiles}(a) -- but rather the 
dominant fluctuation mode. 

\begin{figure}[!t]
\includegraphics[width=0.48\textwidth]{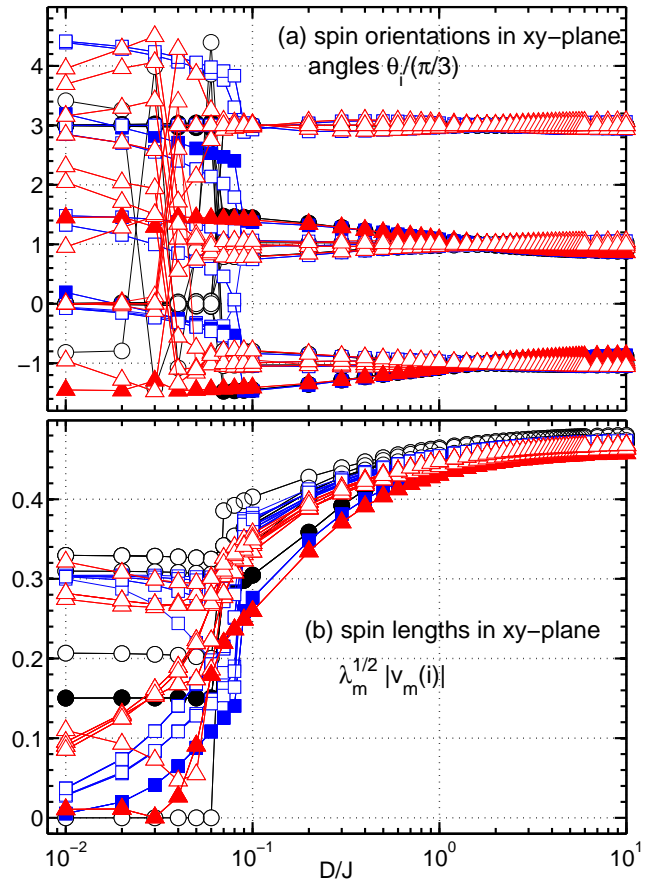}
\caption{(Color online) In-plane spin magnitudes and orientations 
(represented respectively by $\sqrt{\lambda_m} |\bs{v}_m(i)|$ and by the 
phase $\theta_i$ of $\bs{v}_m(i)$) as a function of $D/J$, for clusters 
with $N = 14$ (circles), $20$ (squares), and $26$ (triangles). The data 
corresponding to the four sites closest to the impurity are highlighted 
by filled symbols.}
\label{fig:SpinLength}
\end{figure}

The situation changes dramatically at $D/J \gtrsim 0.1$ 
[Figs.~\ref{fig:CorrDomMode}(c)-(d)], where the system develops 
long-ranged order with the majority of spins participating in the 
$Q = 0$, semiclassical 120$^\circ$ state. The data show clearly that the 
crossover from the dimer-like regime at small $D/J$ to the ordered phase 
at high $D/J$ is smooth, and there are no indications for any other type 
of magnetic order in the intermediate regime $0.06 \lesssim D/J \lesssim 0.1$. 
This suggests once again that the peculiar magnetization pattern found in 
Fig.~\ref{fig:profiles}(b) is most likely to be a fluctuation effect which 
is emphasized close to the critical point. 

Indeed, our analysis of the magnetic correlations allows us to develop 
a qualitative understanding of the situation in this regime. The kagome 
AFM is characterized by a very high density of nearly-degenerate states 
in the ground manifold. The effect of a DM interaction term $D$ is to 
favor certain states and penalize others, but this process is extremely 
sensitive to the shape and size of the cluster used in the calculations. 
Some of our results, such as the discrepancy between Figs.~\ref{fig:profiles}(c) and 
\ref{fig:CorrDomMode}(c), are evidence of significant finite-size effects. 
These effects are expected to, and indeed in 
Figs.~\ref{fig:Cevals} and \ref{fig:scaling} observed to, increase on 
approaching the critical regime around $D/J \sim 0.1$. With such small 
separations between sets of energy levels, the effect of a magnetic field 
applied in this regime can be dramatic: certainly some of the differences 
between Figs.~\ref{fig:profiles} and \ref{fig:CorrDomMode} may be 
ascribed to the finite field in the former. While it remains impossible 
to exclude more exotic types of order for $0.06 \lesssim D/J \lesssim 0.1$, 
whose fingerprints might be difficult or even impossible to discern in 
data obtained from the cluster sizes accessible by exact diagonalization, 
these considerations make systematic our statement that the most probable 
explanation for the tendency of the small-$D/J$ and large-$D/J$ data to 
suggest the presence of an intermediate regime remains in the realm of 
finite-size effects.

The magnetic correlation data also quantify the extent to which the 
four sites closest to the impurity continue to resist the effect of $D$ 
(Fig.~\ref{fig:CorrDomMode}). As already inferred from the behavior of 
the correlation amplitudes (Fig.~\ref{fig:profiles}), these spins do not 
adopt a relative 120$^\circ$ orientation, but remain almost antiparallel 
to each other up to DM interactions $D \sim J$. This direct consequence 
of frustration relief in the two doped triangles also suggests that the 
actual orientation of the ordered component of these spins will continue 
to reflect the behavior of a dimer, and thus will be dictated by $\vec{D}$ 
and $\vec{B}$ as well as by the coupling to the other spins. Only at 
surprisingly high values of $D/J$ does a crossover occur to predominant 
correlation with the other spins of the system, and hence to participation 
in the 120$^\circ$ ordered state.  

This special response of the four sites closest to the impurity is 
demonstrated once again in Fig.~\ref{fig:SpinLength}, which shows the 
orientation and magnitude of the spin components obtained from 
the dominant eigenmode $\bs{v}_m$. Here the spin magnitude at site $i$ is 
represented by $\sqrt{\lambda_m} |\bs{v}_m(i)|$, while the corresponding 
angle is determined from the phase representation of the real and imaginary 
parts of $\bs{v}_m(i)$. The primary feature of Fig.~\ref{fig:SpinLength} is 
that, while both the magnitude and the orientation for the majority of sites 
(open symbols) largely merge together at $D/J \gtrsim 0.06$, the same 
quantities for the four sites closest to the impurity (filled symbols) 
remain different up to $D \sim J$. This feature, present in all three 
clusters, reinforces the picture of the special behavior of the 
frustration-relieved spins. 

\section{Summary and discussion}\label{sec:conclusions}

We have presented an extensive exact-diagonalization study of the 
combined effects of non-magnetic impurities and Dzyaloshinskii-Moriya 
(DM) interactions in the $s = 1/2$ kagome antiferromagnet (AFM). Although 
the clean and purely Heisenberg kagome AFM remains an enigma, being 
apparently non-magnetic and possibly dimer-based, we have shown how 
DM interactions immediately induce a magnetic response, and impurities 
immediately nucleate very strong and somewhat extended patterns of 
dimerization. We have found that the system undergoes a phase transition 
from this magnetically disordered state to the $Q = 0$, three-sublattice 
(120$^\circ$) ordered state when the interaction ratio between the DM and 
AFM terms is $D/J \sim 0.1$. This conclusion is in agreement with the 
results obtained by C\'epas {\it et al.} for the clean kagome 
lattice.\cite{Cepas} 

For $D/J\lesssim 0.06$, there is strong dimerization of the spins next to
each impurity site as a consequence of frustration relief in these two 
triangles, as shown by Dommange {\it et al.}\cite{Dommange} for the $D = 0$ 
case. An oscillating pattern of weakly and strongly dimerized bonds then 
extends some distance from the impurity site. In the presence of DM 
interactions, the applied field induces a magnetic response on top of 
this dimerization pattern, which on the strongly dimerized bonds shows 
the same qualitative behavior as an isolated dimer, namely a dominant 
staggered component directed along $\vec{D} \times \vec{B}$. The induced 
magnetic moments on the sites closest to the vacancy are small as a result 
of their strong dimerization, and we have shown why the magnetization on 
the next-nearest strong bonds is approximately four times larger than this. 
The maximum induced magnetization is thought to occur on these sites or on 
the next (i.e.~next-next-nearest) ``ring'' of more strongly dimerized 
bonds beyond them.\cite{Dommange} Still further from the impurity, the 
local moment decreases again, towards the weak and uniform response 
(by which is meant no staggered component) of the clean kagome system with 
DM interactions. Because this is proportional to $\vec{D} \times (\vec{D} 
\times \vec{B})$, it is much smaller than the staggered response around 
the impurity sites.

The ordered phase established at $D/J \gtrsim 0.1$ shows all the hallmarks 
expected for semiclassical magnetic order. However, we found that the 
spins closest to the impurity remain strongly correlated with each other 
up to surprisingly large values, $D \sim J$, before they begin to 
participate fully in the ordered state. Thus a substantial portion of 
the actual magnetization of these spins remains similar to that of a 
dimer, and is set by $\vec{D}$ and $\vec{B}$. Despite the fact that the 
obvious features of these two phases end at separate values of $D/J$ which 
leave open the possibility of an intermediate phase, between these limits 
we have been unable to find any definite evidence for another type of 
physical behavior. Thus we are forced to the tentative conclusion that 
there is only one true phase transition in the system, and the features 
we have observed for $0.06 < D/J < 0.1$ are most likely to be artifacts 
resulting from the low effective energy scales in this regime and from 
the small system sizes to which our calculations are constrained.

We conclude by discussing the implications of our results for the 
interpretation of experiments performed on ZnCu$_3$(OH)$_6$Cl$_2$.
The most accurate probe of the local magnetic response around doped 
impurities is NMR, and so we focus on the $^{17}$O NMR measurements 
of Olariu {\it et al.}\cite{Olariu} Before comparing the experimental 
results to our calculations, which were performed for a single vacancy 
in small kagome clusters, it is necessary to consider the impurity 
concentrations in the experimental samples. Because these are on the 
order of 5\% for the cleaner samples used in the measurements made to 
date,\cite{Olariu} the average linear spacing between impurities is then 
4-5 Cu--Cu bond spacings: in fact this is rather similar to the lengthscales 
probed in our $N = 20$ and 26 clusters. From our results, it is safe 
to say that there are no Cu spins in the real material which can be 
considered as being far from any impurity sites. 

Our conclusions show that the situation $D/J > 0.1$ may be safely excluded
for herbertsmithite, because a magnetically ordered phase of any type is 
inconsistent not only with the NMR experiments but also with muon spectroscopy 
and inelastic neutron scattering experiments,\cite{muons,Helton} where no 
such order has been found for any temperatures down to 50 mK. By contrast, 
for DM interactions $D/J < 0.1$, one would expect to observe a broad NMR 
signal as a consequence of the considerable number of inequivalent Cu sites 
produced by the partially frozen patterns of dimerization and induced local 
moments around the randomly distributed impurities. This is compatible 
with the shape of the (M) line measured in Ref.~[\onlinecite{Olariu}], while 
the (D) line, which probes only one type of Cu site, would indeed be expected 
to be sharp. Further, our calculations quantify the way in which the DM 
interaction removes the apparent singlet nature of the $D = 0$ ground state, 
leading to the finite lineshift observed in experiment in the limit $T \to 0$. 

To be more specific about the actual strength of $D/J$ in herbertsmithite 
within the disordered regime, we appeal to our results (Sec.~III) for the relative local moments on 
the sites around an impurity. On the qualitative assumption that the $^{17}$O 
line shift should scale with the sum of the moments on the two neighboring 
Cu sites, the 1:4 moment ratio found for the nearest two Cu sites in the 
low-$D$ regime ($D/J \lesssim 0.06$) would be expected to result in a 1:5 
ratio of line shifts. Thus the observed 1:2 ratio\cite{Olariu} suggests 
quite strongly that herbertsmithite falls in the regime $0.06 \lesssim D/J 
\lesssim 0.1$ while the exact nature of the ground state in this region 
remains to be confirmed, it is clear that it is non-magnetic, and that the 
moment ratio between the two types of Cu sites closest to an impurity is 
significantly smaller than at low $D/J$ [Fig.~\ref{fig:SpinLength}(b)]. Such 
a value for $D/J$ is fully consistent with the result $D \simeq 0.08J$ 
obtained by electron spin resonance in Ref.~[\onlinecite{Zorko}].

Finally, we have shown in the disordered regime ($D/J \lesssim 0.1$) that 
sites close to the impurities exhibit strong features characteristic of the 
response of isolated dimers, with those right next to an impurity (the (D) 
line) reflecting a very weak induced magnetization. This behavior could 
explain the suppressed local susceptibility at $T \to 0$ and the enhanced 
spin-relaxation times $T_1$ found\cite{Olariu} for these sites. However, 
to go beyond this qualitative level of agreement, it seems necessary to 
have access to single-crystal data, which would provide a detailed 
understanding of the hyperfine interactions in ZnCu$_3$(OH)$_6$Cl$_2$. 
These are required to investigate the dependence of the magnetic response 
on the field orientation and the line-shift contributions from each Cu 
site as a function of its induced magnetization, information which could 
be interpreted directly within our calculational approach.

\section{Acknowledgments}
We are grateful to O. C\'epas, C. Lhuillier, P. Mendels, and A. Olariu 
for fruitful discussions. This work was supported by the Swiss National 
Science Foundation and by MaNEP.

\appendix
\section{Linear response of the four-site cluster}\label{app:4sites}

\begin{figure}[!t]
\includegraphics[width=0.27\textwidth]{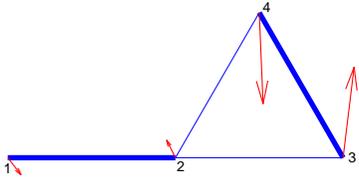}
\caption{(color online) Typical in-plane magnetization pattern (red arrows) 
and spin correlation functions (represented in blue by the bond thickness) 
for the minimal four-site cluster in the small-$D$ regime. Here $D = B = 
J/20$, $\theta=30^\circ$, and the orientations of the DM interactions are 
the same as for sites 1-4 in Fig.~\ref{fig:system}.}
\label{fig:4sites}
\end{figure}

Here we present the details of the magnetic response of the four-site 
cluster shown in Fig.~\ref{fig:4sites}, and demonstrate why this can 
be considered as the minimal cluster illustrating the physics of the 
low-$D$ regime. The Hamiltonian is 
\be
\mc{H} = \mc{H}_1 + \mc{H}_2 + \mc{V} \equiv \mc{H}_0 + \mc{V},
\ee
in which 
\bea
\mc{H}_1 & = & J (\vec{s}_1 \cdot \vec{s}_2 + \vec{s}_3 \cdot \vec{s}_4)
 - \vec{B}\cdot \vec{S},\\
\mc{H}_2 & = & J (\vec{s}_2 \cdot \vec{s}_3 + \vec{s}_2 \cdot \vec{s}_4),\\
\mc{V} & = & \vec{D} \cdot (\vec{s}_2 \! \times \! \vec{s}_1 + \vec{s}_2 
\! \times \! \vec{s}_3 + \vec{s}_3 \! \times \! \vec{s}_4 + \vec{s}_4 \! 
\times \! \vec{s}_2),
\eea
$\vec{S} = \sum_i \vec{s}_i$ is the total spin, $\vec{D} = D \vec{e}_z$, 
and $\vec{B}$ lies in the $xz$-plane, as shown in Fig.~\ref{fig:system}.
Our purpose is to examine the magnetization response of this cluster for 
small fields and to explain why this captures the generic response of the 
corresponding sites around the impurity in Fig.~\ref{fig:profiles}. We 
find numerically that to leading order in $D/J$ there is a staggered 
response along the $y$-axis given by 
\bea
\langle\vec{s}_1\rangle & = & -\langle\vec{s}_2\rangle = \frac{1}{2}\vec{D}
\times\vec{B}\\
\langle\vec{s}_3\rangle & = & -\langle\vec{s}_4\rangle = \frac{21}{8}\vec{D}
\times\vec{B}~,
\eea
i.e.~the magnetization response at bond (1,2) is precisely that expected 
for an isolated dimer while, surprisingly, the response at the bond (3,4) is 
approximately five times greater. We provide a qualitative interpretation of 
this result based on the character of the excitations of $\mc{H}_0$.

Using $\mc{V}$ as a perturbation, the ground state is given to leading 
order by $|\psi \rangle = (1 + \mc{R} \mc{V}) |\psi_0 \rangle$, where 
$|\psi_0 \rangle$ is the ground state of $\mc{H}_0$, $\mc{R} = \frac{Q}
{E_0 - \mc{H}}$ is the corresponding resolvent operator with $Q = 1 - 
|\psi_0 \rangle \langle \psi_0|$, and $E_0 = -3J/2$. The mean value of 
any operator $A$ is then given by 
\be\label{eq:Aavg}
\langle A \rangle = \langle \psi_0 |A \mc{R} \mc{V}| \psi_0 \rangle + 
\textrm{H.c.},
\ee
where we have assumed that $\langle \psi_0 | A | \psi_0 \rangle = 0$.
Because this relation gives the linear response, it may be used to 
obtain the individual contribution of any of the DM terms contained 
in $\mc{V}$ to the magnetization of any spin site.  

For each of the bonds (1,2) and (3,4), it is convenient to use the 
singlet and triplet basis states 
$|s \rangle = (|\uparrow\downarrow\rangle - |\downarrow\uparrow\rangle)/
\sqrt{2}$,
$|t_{1}\rangle=|\uparrow\uparrow\rangle$, 
$|t_0 \rangle = (|\uparrow\downarrow\rangle + |\downarrow\uparrow\rangle)/
\sqrt{2}$, and 
$|t_{-1}\rangle=|\downarrow\downarrow\rangle$.
For $B < J/2$, the ground state $|\psi_0 \rangle$ of $\mc{H}_0$ is the 
product state of the two singlets,
\be
|\psi_0\rangle = |s\rangle_{12} \otimes |s\rangle_{34}.
\ee
Also applicable are the relations\cite{Miyahara2} 
\bea
&&\mc{H}_2 |s\rangle_{12}\otimes|s\rangle_{34} = 0\\
&&\mc{H}_2 |t_m\rangle_{12}\otimes|s\rangle_{34} = 0, \;\; \forall m=0,\pm1,
\eea 
which specify that the state $|t_m \rangle_{12} \otimes |s \rangle_{34}$ 
is also an eigenstate of $\mc{H}_0$. Thus the local excitation of a triplet 
on bond (1,2), for example by acting with the term $\vec{D} \cdot \vec{s}_2 
\times \vec{s}_1$ (or the operators $A = s_{1,2}^y$) on $|\psi_0\rangle$, 
gives 
\be
\mc{R} |t_m\rangle_{12}\otimes |s\rangle_{34} = \frac{-1}{1+m B} |t_m 
\rangle_{12}\otimes |s\rangle_{34},
\ee
and the triplet remains localized. 

By constrast, a triplet on bond (3,4) will not remain localized, because  $|s 
\rangle_{12} \otimes |t_m \rangle_{34}$ is not an eigenstate of $\mc{H}_0$. 
With these considerations and by using Eq.~(\ref{eq:Aavg}), it is easy to 
show that of all the DM terms contained in $\mc{V}$, only $\vec{D} \cdot 
\vec{s}_2 \times \vec{s}_1$ contributes to $\langle \vec{s}_{1,2} \rangle$, 
and this explains in turn why the response at bond (1,2) is that of an 
isolated dimer. The staggered magnetization at bond (3,4), on the other 
hand, is driven by all the terms of $\mc{V}$ except $\vec{D} \cdot \vec{s}_2 
\times \vec{s}_1$, which explains why this is different from the response of 
bond (1,2). More generally, we see that although the ground state of $\mc{H}_0$ 
is a product of two dimers, the different character of the local excitations 
leads to quantitatively very different responses. This completes the 
qualitative interpretation, summarized in Sec.~\ref{sec:local}, for the 
different magnetization responses of the sites around the impurity in 
Fig.~\ref{fig:profiles} 

\section{Interpretation of magnetic correlations}\label{app:Cij} 

Here we provide a more detailed presentation of the natural orbital method 
used in Sec.~\ref{sec:correlations} for the study of in-plane magnetic 
correlations. We consider the ground state $|\Psi\rangle$ of the Hamiltonian 
given in Eq.~(\ref{eq:hamiltonian}) and express the in-plane magnetic 
correlations by the matrix $\vec{C}$ of Eq.~(\ref{eq:Cij}). Let us denote 
by $\{\lambda_a, \bs{v}_a\}$ the set of eigenvalues and normalized 
eigenvectors of $\vec{C}$, i.e.~$\vec{C} \cdot \bs{v}_a = \lambda_a 
\bs{v}_a$. 
In addition to being Hermitian, the correlation matrix $\vec{C}$ is also 
positive semi-definite, because for any normalized vector $\bs{v}$ one has 
$\bs{v}^\dagger \cdot \vec{C} \cdot \bs{v} = \| \mc{M}_{\bs{v}} |\Psi 
\rangle \|^2 \ge 0$, where 
\be\label{eq:OP}
\mc{M}_{\bs{v}} \equiv \sum_i \bs{v}(i) s_i^-
\ee
defines a macroscopic magnetic mode. Thus $\lambda_a \geq 0$ for all $a$. 
Further, because $s^+ s^- = 1/2 + s^z$ for $s = 1/2$, $\text{Tr} \vec{C}
 = \sum_a \lambda_a = N/2 + \langle S^z \rangle$. We note also that the 
eigenvalues $\lambda_a$ give the fluctuations of the matrix $\mc{M}_{a}$
(specified by Eq.~(\ref{eq:OP}) with $\bs{v} = \bs{v}_a$), because
\be\label{eq:OO1}
\langle \Psi| \mc{M}_{a}^\dagger \mc{M}_{b} |\Psi\rangle = \lambda_a 
\delta_{a b}.
\ee

In order to discuss magnetic order, we consider the thermodynamic limit 
and address the question of what is required to ensure long-ranged in-plane 
magnetic correlations, i.e. 
\be\label{eq:LRO}
\lim_{|i-j|\to\infty}C_{ij} \ne 0.
\ee 
To this end, it is convenient to express $\vec{C}$ in terms of its spectral 
decomposition, 
\be\label{eq:Cij1}
C_{ij} = \sum_a \lambda_a \bs{v}_a(i) \bs{v}_a^*(j).
\ee
Equations (\ref{eq:LRO}) and (\ref{eq:Cij1}) mean that there should exist 
at least one eigenstate $\bs{v}_m$ with a nonzero amplitude $\bs{v}_m(i)$ 
for all $i$. However, $\bs{v}$ is a quantity normalized to one by an 
overall factor of $1/\sqrt{N}$, which from Eq.~(\ref{eq:LRO}) means that 
the corresponding eigenvalue is macroscopically large, $\lambda_m \propto N$.
If only one (the maximum) eigenvalue has this property, it is safe to 
replace Eq.~(\ref{eq:Cij1}) with 
\be\label{eq:Cij2}
C_{ij} \simeq \lambda_m \bs{v}_m(i) \bs{v}_m^*(j),
\ee
because the neglected terms are negligibly small for $N \to \infty$. 
This special, separable form of the correlation matrix in an ordered state
allows one to identify the relevant local order parameter. Indeed, in an 
explicitly symmetry-broken (coherent) state we may replace $\langle s_i^+ 
s_j^-\rangle$ by $\langle s_i^+\rangle\langle s_j^-\rangle$, and hence 
make the identification
\be\label{eq:lop}
\langle s_i^+ \rangle = \langle s_i^x \rangle + i \langle s_i^y \rangle = 
\sqrt{\lambda_m} \bs{v}_m(i).
\ee
We remark here that these in-plane spin components are fixed only up to 
a global $\mathsf{U}(1)$ rotation, due to the fact that the eigenstate 
$\bs{v}_m$ is only specified up to a global phase. The corresponding 
macroscopic order parameter is given by $\mc{M}_{m}$, because [using 
Eq.~(\ref{eq:OO1})]
\be\label{eq:OO2}
\langle \Psi| \mc{M}_{m}^\dagger \mc{M}_{m} |\Psi \rangle = \lambda_m 
\propto N,
\ee
and thus the response to a conjugate field that couples directly to 
$\mc{M}_{m}$ diverges in the thermodynamic limit.

Finally we should mention the analogy of the discussion presented in this 
appendix to Bose-Einstein condensation, and in particular to the case of 
cold-atom systems confined in a harmonic trap (see for example 
Ref.~[\onlinecite{DuBois}]). The correlation matrix in such a system is 
the one-body density matrix, $\rho_1(\vec{x},\vec{x}')$, which measures 
the coherence between different parts of the system. The eigenstates of 
$\rho_1$ are termed ``the natural orbitals'', and the eigenvalues give the 
relative occupation probability of these orbitals. As in our situation, 
off-diagonal, long-range order is signaled by the fact that one of the 
eigenvalues of $\rho_1$ becomes macroscopically large.\cite{Penrose,Yang}
The dominant eigenstate $\bs{v}_m$ of a magnetic system is thus analogous 
to the condensate wavefunction in a superfluid.


\begin{thebibliography}{99}
\bibitem{factor2} We note here that the the extrapolated values of the (unrescaled) squared moments reported for the clean kagome system in Ref.~[\onlinecite{Cepas}] are smaller than the true values by a factor of two. 
The origin of this factor lies in the choice of the $x$-components of the spins for studying the 
correlations: The dominant eigenvalue of these correlations is twofold degenerate 
and corresponds to two modes with the same magnetization pattern in the $x$-direction but with different chirality. By contrast, the correlation matrix $\vec{C}$ defined in Eq.~(\ref{eq:Cij})
is able to select the state with the right chirality, i.e. the one favored by the DM interactions.
\end{thebibliography}
\end{document}